\documentstyle[graphicx]{./mn}

\newif\ifAMStwofonts

\def\mnras{MNRAS}
\def\apj{ApJ}

\def\aap{A\&A}

\def\nat{Nat}

\def\iaucirc{IAUC}

\ifoldfss
  \ifCUPmtlplainloaded \else
    \NewTextAlphabet{textbfit} {cmbxti10} {}
    \NewTextAlphabet{textbfss} {cmssbx10} {}
    \NewMathAlphabet{mathbfit} {cmbxti10} {} 
    \NewMathAlphabet{mathbfss} {cmssbx10} {} 
  \fi
  \ifAMStwofonts
    \ifCUPmtlplainloaded \else
      \NewSymbolFont{upmath} {eurm10}
      \NewSymbolFont{AMSa} {msam10}
      \NewMathSymbol{\upi}     {0}{upmath}{19}
      \NewMathSymbol{\umu}     {0}{upmath}{16}
      \NewMathSymbol{\upartial}{0}{upmath}{40}
      \NewMathSymbol{\leqslant}{3}{AMSa}{36}
      \NewMathSymbol{\geqslant}{3}{AMSa}{3E}

      \let\leq=\leqslant 
       
    \fi
  \fi
\fi 

\ifnfssone
  \newmathalphabet{\mathit}
  \addtoversion{normal}{\mathit}{cmr}{m}{it}
  \addtoversion{bold}{\mathit}{cmr}{bx}{it}
  \newmathalphabet{\mathbfit} 
  \addtoversion{normal}{\mathbfit}{cmr}{bx}{it}
  \addtoversion{bold}{\mathbfit}{cmr}{bx}{it}
  \newmathalphabet{\mathbfss} 
  \addtoversion{normal}{\mathbfss}{cmss}{bx}{n}
  \addtoversion{bold}{\mathbfss}{cmss}{bx}{n}
  \ifAMStwofonts
    \ifCUPmtlplainloaded \else
      \UseAMStwoboldmath
      \makeatletter
      \new@mathgroup\upmath@group
      \define@mathgroup\mv@normal\upmath@group{eur}{m}{n}
      \define@mathgroup\mv@bold\upmath@group{eur}{b}{n}
      \edef\UPM{\hexnumber\upmath@group}
      \new@mathgroup\amsa@group
      \define@mathgroup\mv@normal\amsa@group{msa}{m}{n}
      \define@mathgroup\mv@bold\amsa@group{msa}{m}{n}
      \edef\AMSa{\hexnumber\amsa@group}
      \makeatother
      \mathchardef\upi="0\UPM19
      \mathchardef\umu="0\UPM16
      \mathchardef\upartial="0\UPM40
      \mathchardef\leqslant="3\AMSa36
      \mathchardef\geqslant="3\AMSa3E

      \let\leq=\leqslant 

    \fi
  \fi
\fi 

\ifnfsstwo
  \DeclareMathAlphabet{\mathbfit}{OT1}{cmr}{bx}{it}
  \SetMathAlphabet\mathbfit{bold}{OT1}{cmr}{bx}{it}
  \DeclareMathAlphabet{\mathbfss}{OT1}{cmss}{bx}{n}
  \SetMathAlphabet\mathbfss{bold}{OT1}{cmss}{bx}{n}
  \ifAMStwofonts
    \ifCUPmtlplainloaded \else
      \DeclareSymbolFont{UPM}{U}{eur}{m}{n}
      \SetSymbolFont{UPM}{bold}{U}{eur}{b}{n}
      \DeclareSymbolFont{AMSa}{U}{msa}{m}{n}
      \DeclareMathSymbol{\upi}{0}{UPM}{"19}
      \DeclareMathSymbol{\umu}{0}{UPM}{"16}
      \DeclareMathSymbol{\upartial}{0}{UPM}{"40}
      \DeclareMathSymbol{\leqslant}{3}{AMSa}{"36}
      \DeclareMathSymbol{\geqslant}{3}{AMSa}{"3E}

      \let\leq=\leqslant 

    \fi
  \fi
\fi 

\ifCUPmtlplainloaded \else
  \ifAMStwofonts \else 
    \def\upi{\pi}
    \def\umu{\mu}
    \def\upartial{\partial}
  \fi
\fi

\title{The radio remnant of SN1993J: an instrumental explanation for the evolving complex structure}
\author[Heywood et al]
{Ian~Heywood$^{1}$, Katherine~M.~Blundell$^{1}$, Hans-Rainer Kl\"{o}ckner$^{1}$, Anthony~J.~Beasley$^{2}$\\$^{1}$~University of Oxford Astrophysics, Keble Road, Oxford, OX1 3RH, UK\\$^{2}$~NEON, Inc., 3223 Arapahoe Ave., Suite 210, Boulder, CO 80303, USA}
\date{Accepted 2008 month 10. Received 200X month XX. Preprint version for astro-ph}
\pagerange{\pageref{firstpage}--\pageref{lastpage}}
\pubyear{200X}

\def\LaTeX{L\kern-.36em\raise.3ex\hbox{a}\kern-.15em
    T\kern-.1667em\lower.7ex\hbox{E}\kern-.125emX}

\begin{document}

\label{firstpage}

\maketitle

\begin{abstract}

We present simulated images of Supernova 1993J at 
8.4~GHz using Very Long Baseline Interferometry (VLBI) techniques. A spherically symmetric source model is convolved with realistic $uv$-plane distributions, together with standard
imaging procedures, to assess the extent of instrumental effects on the recovered brightness distribution. In order to facilitate 
direct comparisons between the simulations and published 
VLBI images of SN1993J, the observed $uv$-coverage is determined from actual VLBI observations made in the years following its discovery. 

The underlying source model only exhibits radial variation in its density profile, with no azimuthal dependence
and, even though this model is morphologically simple, the simulated VLBI observations qualitatively reproduce many of the azimuthal features of the reported VLBI 
observations, such as appearance
and evolution of complex azimuthal structure and apparent rotation of the shell. We demonstrate that such features are inexorably coupled to the $uv$-plane sampling.

The brightness contrast between the peaks and the 
surrounding shell material are not as prominent in the simulations (which of course assume no antenna- or baseline-based 
amplitude or phase errors, meaning no self-calibration procedures will have incorporated any such features in models). It is conclusive that 
incomplete $uv$-plane sampling has a drastic effect on the final images for observations of this nature. Difference imaging reveals residual emission
up to the 8$\sigma$ level.
Extreme care should be taken when using interferometric observations to directly infer the structure of objects such 
as supernovae.

\end{abstract}


\section{Introduction}

Supernova 1993J (RA~=~09h~55m~24.7747s, $\delta$~=~+69$^{\circ}$~01'~13.7031'') was discovered on 
28 March 1993 in the nearby spiral galaxy M81 (Ripero et al., 1993) and became the 
brightest radio supernova ever observed. Its brightness and high Declination have allowed this object to be studied in great detail at radio wavelengths. 

A shell-like radio structure associated with SN1993J was resolved using VLBI techniques as early as 
239 days after the supernova explosion (Marcaide et al., 1995). An extensive VLBI monitoring program then followed which 
tracked the expansion of the shell over seven years at three frequencies (Bietenholz, Bartel \& Rupen, 2001; 
Bartel et al., 2002; Bietenholz, Bartel \& Rupen, 2003). These observations yielded a wealth of information 
about the supernova, including accurate determination of the explosion centre, measurement of the deceleration 
of the ejecta and radio light curves and spectra. Images of the resolved shell allowed direct monitoring of 
its development, from its initial barely resolved state, through expansion into partial shell structures, to 
the complex structure apparently present in the final epochs 3000+ days later. 

Behaviour such as apparent rotation of features in the shell in the plane of the sky are seen between epochs of VLBI 
observations, as are phenomena such as coherent structures dividing into a series of peaks which apparently 
change in azimuthal position as the remnant expands. Bietenholz, Bartel and Rupen (2003) discuss possible physical mechanisms 
which could explain the structure in the shell, including Rayleigh-Taylor cooling instabilities, evolving magnetic 
field structure and interactions with inhomogeneous circumstellar material. 

This paper presents a new investigation of this intriguing evolution of the brightness distribution in the expanding 
shell. We explore whether the development of apparently complex structure 
may be due to incomplete sampling of the $uv$ plane, which is problematic especially for sparse interferometer 
arrays such as VLBI networks. Similar effects had also been a cause for concern when interpreting 6-cm MERLIN images 
of nova Cassiopeiae 1995 (Heywood et al. 2005; Heywood \& O'Brien, 2007) where the apparently complex development 
of the radio emission could be accounted for by a spherically symmetric shell model and sparse sampling of the $uv$-plane. 
With this in mind we simulated 8.4-GHz VLBI observations of the radio 
shell of SN1993J at nine epochs, for comparison with the genuine VLBI images. 

Further discussion regarding the non-uniqueness of deconvolved radio images (particularly due to the CLEAN
deconvolution method) may be found in e.g. Cornwell, Braun and Briggs (1999) and Briggs (1995). The latter
contains a particularly relevant case study of the radio shell of SN1987A whereby different deconvolution methods yield subtle
variations in apparent structure.

\section{Simulations}

Creation of the simulated VLBI images requires two components. First, a model of the assumed source brightness 
distribution is required. This model should represent how an ideal instrument 
would image the source at a given frequency. The second component of the simulation is a $uv$-plane sampling 
function which accurately represents that of the real instrument.

\subsection{Model brightness distributions}

\begin{figure*}
\begin{center}
\setlength{\unitlength}{1cm}
\begin{picture}(16,11.0)
\put(-0.5,0.0){\includegraphics{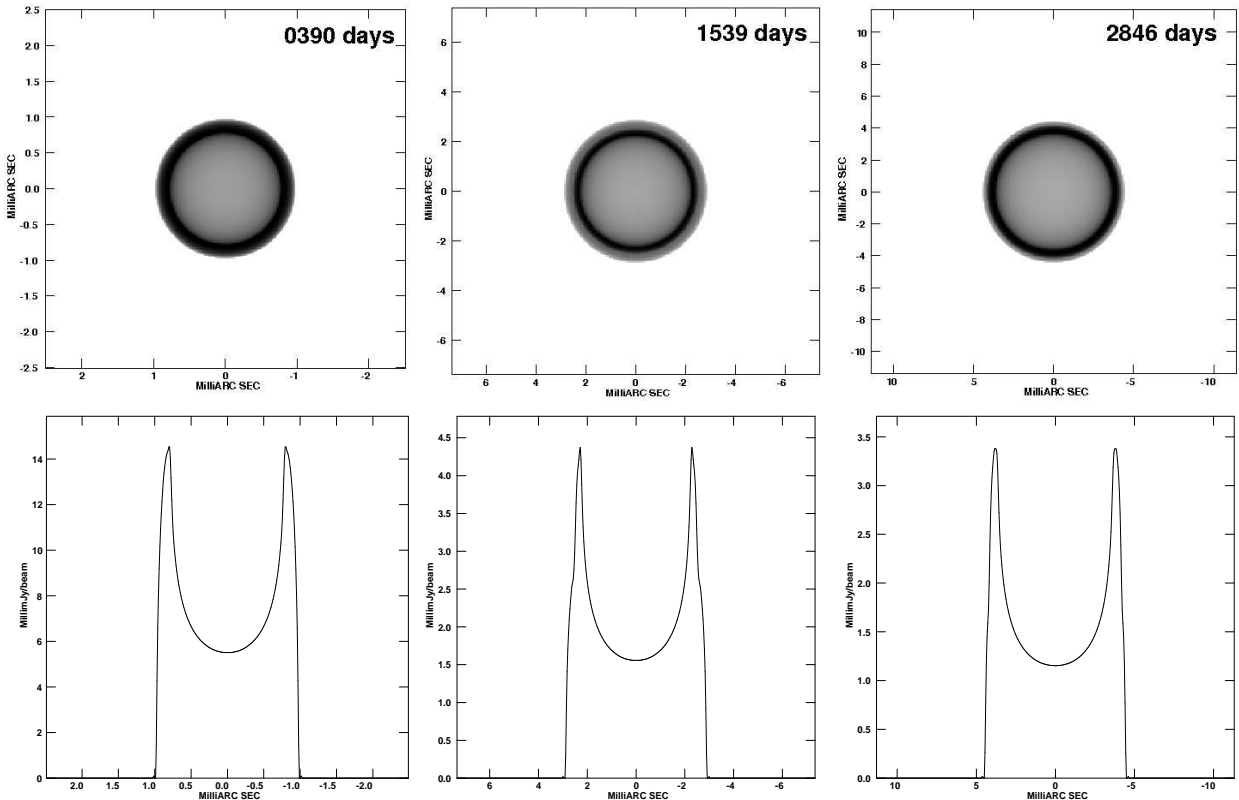}}
\end{picture}
\caption{\label{fig:inputmodels} Model radio brightness distributions (upper row) and their corresponding brightness profiles (lower row) for three epochs of the parametrized synchrotron model for SN1993J (Mioduszewski et al., 2001). 
These models are not self-similar, having radial evolution, but they are entirely azimuthally symmetric.}
\end{center}
\end{figure*}

A synchrotron shell model is used as the basis for our simulated radio images. The expansion of the shell in 
this model is not self-similar because the radial density profile is time dependent as the physical shell boundary 
expands, and reverse shocks propagate back through the ejecta. The model is fully spherically symmetric however, 
and the density variations are purely radial with no azimuthal component. A detailed description of this model is provided by Mioduszewski 
et al. (2001). Simulated radio images were used with ages that closely matched the epochs of the VLBI observations, and were evenly 
spaced throughout the evolution of SN1993J. Mioduszewski et al. (2001) also parametrized and fitted their models to the radio
observations to give consistent angular sizes and flux density values, and kindly allowed us to use their models in the
investigations presented here. Figure \ref{fig:inputmodels} shows three epochs of the simulated radio images and their corresponding
flux density profiles. 

\subsection{Recreating the $uv$ coverage}

In addition to the model images being chosen to coincide as best as possible with the genuine VLBI 
observations the model array for each epoch is defined to be the same as those used during the SN1993J VLBI campaign. 
Only stations which were used during an actual observation are included for the corresponding simulation. 

The AIPS task UVCON is used to generate the model $uv$-data sets. By examining VLBA archive files, the 
start and end time for any observation are used to determine the total duration. Using the relevant model 
image and an antenna file describing the stations present for a specific observation, UVCON 
is executed, resulting in a track length equivalent to this total duration. In the case of the real observation, this track would 
encompass all sources and frequencies. 

In order to reduce the $uv$-data set such that it only includes visibilities related to 
SN1993J at 8.4~GHz three flag tables are applied, documenting the frequency switching 
information, the SN1993J scan times, and the automated flagging that occurred at the receivers.

These flag tables are constructed from archival data files, and converted to a format that can be directly read by the UVFLG task in AIPS. 
Bad data that are manually removed from the true VLBI observations in the pre-calibration stage cannot be accounted 
for as the simulations are not prone to problems such as radio frequency interference. This means that in 
each case the simulations are over-estimating the $uv$-plane sampling, and that our analysis errs on the side of being conservative.

Multi-frequency interferometric observations often employ a technique to improve $uv$-plane coverage, whereby 
the observing frequency is switched periodically. Consider an 18-hour track at three frequencies. Rather 
than observing at each frequency in three successive six-hour blocks the observing frequency is switched, for example 
every 20 minutes. The visibilities for a given frequency are therefore more spread out on the $uv$-plane, 
facilitating the sampling of more diverse Fourier components and reducing linear biases. Determining this 
frequency switching information allows us to flag out scans which do not occur at 8.4~GHz. 

Once the simulated $uv$-tracks are consistent only with the scans at 8.4~GHz we flag out times where the calibration sources are being observed.
From the archive data we can obtain the times when SN1993J was being observed and use this information to retain only these time ranges in the simulation.
AIPS-readable flag tables describing data that were flagged at the time of the observation (e.g. due to slewing or pointing errors, source changes 
etc.) are also available, and this is the final flag table applied to the simulated data set.

Note also that a 15$^{\circ}$ elevation cut off for each antenna is also imposed on the simulation. Data are routinely flagged
below this elevation because system temperatures rise as the antenna points closer to the horizon, and local
geographical features can often obscure the line of sight. For a source at the Declination of SN1993J this condition only slightly
affects the longest baselines in the VLBA network.

The UVCON task also adds random Gaussian noise to the simulated data set, the level of which is governed by the parameters of each
antenna in the array, such as the antenna efficiency and system temperature.

\subsection{Cleaning and contouring}

The simulated $uv$-data sets were imaged using the CLEAN algorithm (Clark, 1980) using the AIPS task IMAGR. For each simulation the circular restoring 
beam used is consistent with that of the nearest epoch of the VLBI observations, as reported by Bietenholz, Bartel and 
Rupen (2003). Briggs (1995) weighting was utilised with the ROBUST parameter in IMAGR set to zero. For the nine simulated maps presented in Figure \ref{fig:ninemaps}
the data were reweighted
by the square roots of the data weights using the UVWTFN parameter. This is a common technique for VLBI observations and is consistent with the
imaging methods of Bietenholz, Bartel and Rupen (2003). Deconvolution 
was carried out interactively and the clean cycle terminated when it was judged that all substantial flux 
has been removed. 

Since no antenna-based or baseline-based amplitude or phase errors are included in our simulations, 
this means no self-calibration procedures will be responsible for spuriously enhancing features in our simulations. Our simulations will only
underestimate the extent to which small compact features could become \emph{reinforced} in self-calibration and cleaning cycles.

Contouring of the simulated images is also carried out in a way that is consistent with the images presented by Bietenholz, Bartel 
and Rupen (2003). The greyscale in each image runs from 8\% to 100\% of the peak brightness for epochs below 
1253 days and from 16\% to 100\% for epochs above. Contour levels are 1\%, 2\%, 4\%, 8\%, 16\%, 32\%, 45\%, 
64\% and 90\% of the peak brightness in each case, starting with the first contour level that is brighter than three times the root 
mean square background brightness levels (3$\sigma$). Presenting the simulated images in a consistent way better facilitates comparison 
with the real observations.

\section{Simulated observations}

In this section we first present the nine simulated radio images which best recreate the genuine VLBI observations of SN1993J.
We discuss aspects of the real VLBI observations which differ from our simulations. Also presented in this section are residual images
created by differencing the simulated radio image from a convolved input model.

\subsection{Simulated VLBI Images}

\begin{figure*}
\begin{center}
\setlength{\unitlength}{1cm}
\begin{picture}(10,17.5)
\put(-3.3,0){\includegraphics{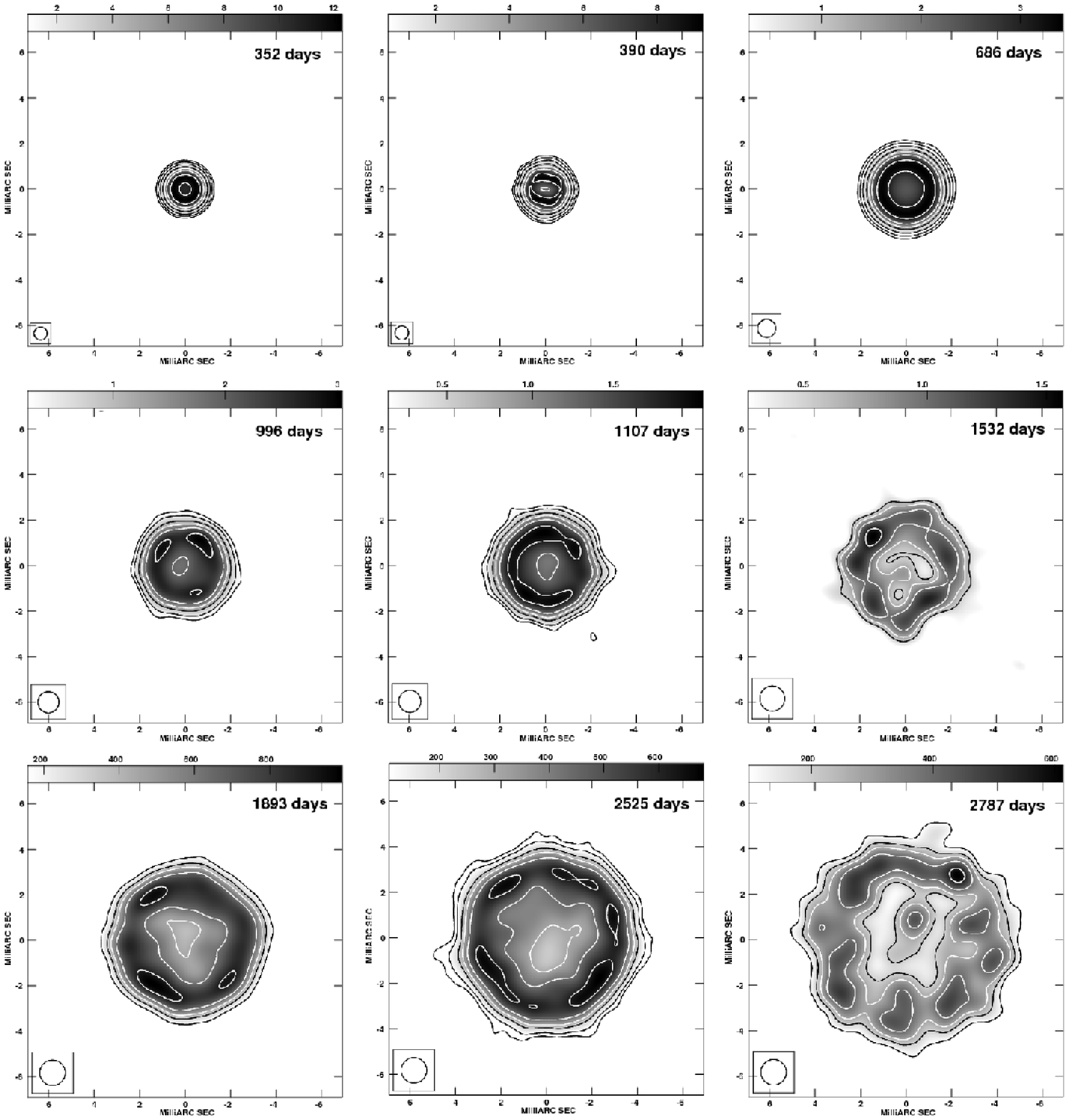}}
\end{picture}
\caption{Simulated VLBI images at nine successive epochs. Contour levels are 1\%, 2\%, 4\%, 8\%, 16\%, 32\%, 45\%, 
64\% and 90\% of the peak brightness starting with the first contour level above 3$\sigma$. The greyscale is from 8\% to 100\% of the peak brightness for 
$t$~$\leq$~1253 days, and from 16\% to 100\% for later epochs. The greyscale pixel values above the images are in units of mJy/beam, except for the final three epochs which 
are in $\mu$Jy/beam.\label{fig:ninemaps}}
\end{center}
\end{figure*}

Figure \ref{fig:ninemaps} shows the resulting images from nine simulated VLBI observations of SN1993J. 
A summary of the relevant image parameters can be found in Table \ref{tab:images}. The total shell flux density values
for the simulated VLBI maps are measured by using the AIPS task TVSTAT to define the region of emission. 

\begin{table*}
\centering
\begin{minipage}{170mm}
\caption{Summary of the image parameters for the input model and the simulations. \label{tab:images}}
\begin{tabular}{lllllll}
\hline
Observation  &\multicolumn{2}{c}{Input model}      & \multicolumn{3}{c}{Simulated VLBI images}                      & Restoring \\
time (days)  &Time (days) & Flux density (mJy) & Flux density (mJy) & Peak (mJy/beam)     & 3$\sigma$ (mJy/beam)& beam (mas) \\ \hline
352          &321         & 86.28              & 85.56              & 12.24               & 0.087  & 0.57   \\
390          &390         & 71.86              & 70.74              & 9.21                & 0.143  & 0.60   \\
686          &707         & 38.79              & 38.34              & 3.43                & 0.038  & 0.80   \\
996          &867         & 31.06              & 30.82              & 3.04                & 0.213  & 0.93   \\
1107         &1062        & 25.57              & 24.86              & 1.99                & 0.148  & 0.97   \\
1532         &1539        & 19.26              & 17.01              & 1.56                & 0.276  & 1.10   \\
1893         &1939        & 18.32              & 17.91              & 0.99                & 0.066  & 1.12   \\
2525         &2351        & 16.84              & 16.38              & 0.68                & 0.037  & 1.12   \\
2787         &2846        & 15.15              & 14.81              & 0.61                & 0.067  & 1.12   \\
\hline
\end{tabular}
\end{minipage}
\end{table*}

\subsection{Evolution of the simulated radio images}

The evolution of the simulated VLBI images as shown in Figure \ref{fig:ninemaps} has much in common with that of the genuine VLBI observations of 
Bietenholz, Bartel and Rupen (2003), with the major difference occuring in the early stages. Specifically, in both the 
case of the real observations and the simulations the emission is intially barely resolved, and as the shell
expands the real observations begin to take on a partial shell structure, which is only hinted at in the 
simulated image corresponding to day 996. While in the VLBI observations this partial shell structure expands 
and appears to rotate, the simulated images deviate very little from circularity until day 1532 when a 
fragmented shell develops.

The fragmented shell continues to expand, giving the impression of ridge structures dividing into peaks and 
undergoing apparent rotation. The simulation with the largest angular extent is remarkably similar to the VLBI 
observation at day 2787. 

Although the ring-like structure upon which the peaks are situated seems far more coherent in the simulations 
they are still qualitatively very similar to the VLBI images. Given that the highest contour represents 90\% 
of the peak brightness and the next lowest one is at the 64\% level these brightness enhancements represent a 
significant deviation from circularity, which given the nature of our input model can only be a purely 
instrumental effect. If the structure apparent in the VLBI images of SN1993J represents genuine clumpiness 
of the ejecta then at best a degeneracy exists between structures of that nature and ring-like objects.
 
\subsection{How the model deviates from the reality}

Our simulated images can qualitatively reproduce the effects apparent in the VLBI observations of SN1993J, however some differences are present
and this section suggests some possible reasons why such differences may arise. Deviations from a perfect reconstruction of the SN1993J VLBI observations
will occur either in the $uv$-plane model or the radio brightness distribution model. 

An obvious reason why the $uv$-distributions will deviate is that our simulations are not prone to radio frequency interference. Any contamination of the real
data by such inteference would be flagged out, therefore in this respect our simulations are over-estimating the sampling of the $uv$ plane. In addition to this, as mentioned elsewhere,
the simulations are not prone to calibration errors which may result in spurious features that are reinforced due to self-calibration.

With respect to the model brightness distribrutions employed, it may well be the case that the radio remnant of SN1993J is \emph{not} spherically symmetric
with solely radial structure variations. Indeed, imaging nearby supernova remnants at many wavelengths suggests that complex morphologies are common. However, should it be the case
that SN1993J deviates from sphericity as much as the VLBI images suggest, our results demonstrate that a degeneracy exists between such complex structures and simple morphologies, 
with VLBI observations not being able to distinguish between the two.

For each element in our simulated array we also assume a fixed system temperature ($T_{\rm sys}$). However, in a real observation $T_{\rm sys}$ will increase (e.g. as the antenna elevation decreases)
and will affect the noise levels in the observations more than is accounted for in our simulations which are not prone to these effects. Note that as the source becomes fainter
and the signal to noise level decreases, the noise in the image exacerbates the apparent fragmented shell.

\subsection{Residual structures}

\begin{figure*}
\begin{center}
\setlength{\unitlength}{1cm}
\begin{picture}(10,17.5)
\put(-3.3,0){\includegraphics{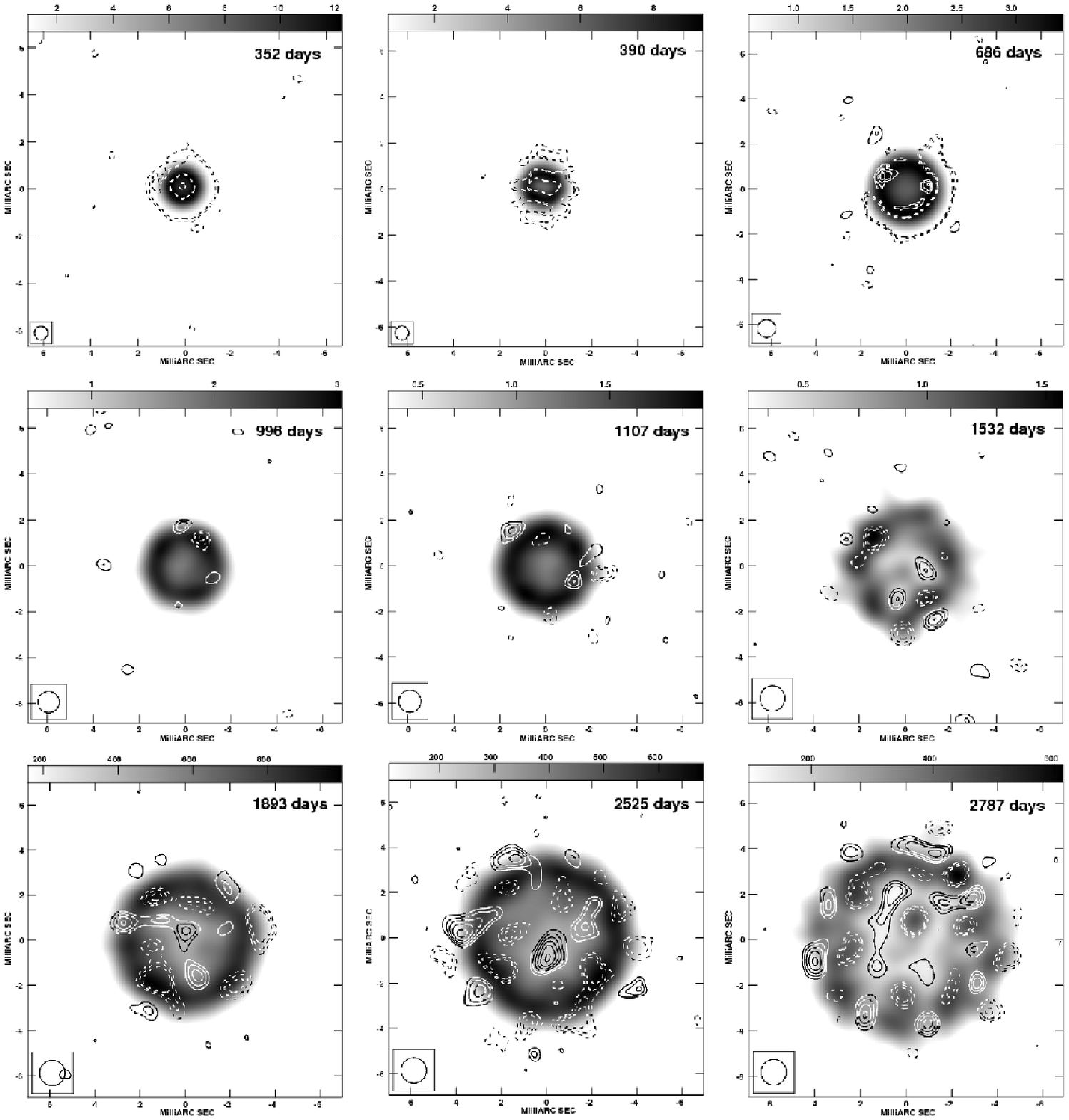}}
\end{picture}
\caption{Residual maps created by subtracting the simulated VLBI image (shown in Figure \ref{fig:ninemaps}) from an appropriately convolved input model. Contours show the difference map,
superimposed on a greyscale plot of the simulated VLBI image. Contour levels are (-5, -4, -3, 3, 4, 5, 6, 7, 8)
times the rms background level ($\sigma$) for each specific image. The greyscale runs from 8\% to 100\% (t~$\leq$~1253 days) and 16\% to 100\% (thereafter) 
of the peak brightness in the simulated image.
This is shown in the scale above each image, the units of which are mJy / beam. 
The $\sigma$ values are consistent with the 3$\sigma$ values presented in Table \ref{tab:images}. \label{fig:differencemaps}}
\end{center}
\end{figure*}

Figure \ref{fig:differencemaps} shows the simulated VLBI images (presented in Figure \ref{fig:ninemaps}) subtracted from the input model.
To facilitate this comparison the input models are convolved with circular Gaussians with sizes equal to those of the restoring beams used to produce
the simulated images. The restoring beams for each epoch are listed in Table \ref{tab:images}. The images are carefully aligned to give identical central pixel values and pixel increments.

The residuals from the subtraction are contoured and superimposed over the simulated VLBI image, with contour levels of (-5, -4, -3, 3, 4, 5, 6, 7, 8) $\times$
$\sigma$. The underlying greyscales are the same as those presented in Figure \ref{fig:ninemaps} and the values of $\sigma$ are in accordance with 
those presented in Table \ref{tab:images}.

By contouring the residuals above the 3$\sigma$ level the structure that is revealed can be considered to represent genuine artifacts arising due to the $uv$-plane sampling, 
the deconvolution process, or both. Note how the brightest features in the simulated image often correspond to negative regions in the residual map. 
The effects are less pronounced in the earlier epochs which exhibit higher signal to noise levels for the source, as well as
having more compact angular sizes. Even in the final and faintest epoch there is residual structure at the 7$\sigma$ level. 

We quantify the level of the residual structure by measuring the rms of the signal over the region occupied by the shell. This is presented in Table \ref{tab:residuals}
both as an absolute value and in units of the background rms ($\sigma$). The peak and minimum flux densities are also presented in order to represent the extremes
of the image contamination. 

\begin{table*}
\centering
\begin{minipage}{110mm}
\caption{Peak and root mean square flux levels over the emitting region of the shell. The latter is expressed both as an absolute value and 
in terms of the background noise ($\sigma$; 3$\sigma$ values are listed in Table \ref{tab:images}.) \label{tab:residuals}.}
\begin{tabular}{lllll}\hline
Observation  & Peak flux  & Minimum flux & Residual RMS & Residual RMS \\ 
time (days)  & (mJy/beam) & (mJy/beam)    & (mJy/beam)   & $/$ $\sigma$    \\ \hline
352          & 0.392      & -0.711       & 0.194        & 6.7\\
390          & 0.937      & -0.989       & 0.233        & 4.9\\
686          & 0.073      & -0.235       & 0.056        & 4.4\\
996          & 0.316      & -0.346       & 0.089        & 1.3\\
1107         & 0.233      & -0.255       & 0.073        & 1.5\\
1532         & 0.485      & -0.527       & 0.142        & 1.5\\
1893         & 0.139      & -0.167       & 0.047        & 2.2\\
2525         & 0.108      & -0.109       & 0.033        & 2.7\\
2787         & 0.177      & -0.232       & 0.061        & 2.7\\ \hline
\end{tabular}
\end{minipage}
\end{table*}

\section{Illustrative simulations}

The first part of this section presents simplified simulated observations to demonstrate the coupling between
the $uv$-plane sampling and the resulting image. The second subsection takes a more realistic simulation
and shows how subtle changes to the visibility data (e.g. antenna flagging) affect the image.

\subsection{The dependence of azimuthal structure on hour angle}

\begin{figure*}
\centering
\includegraphics[width= 0.76 \columnwidth]{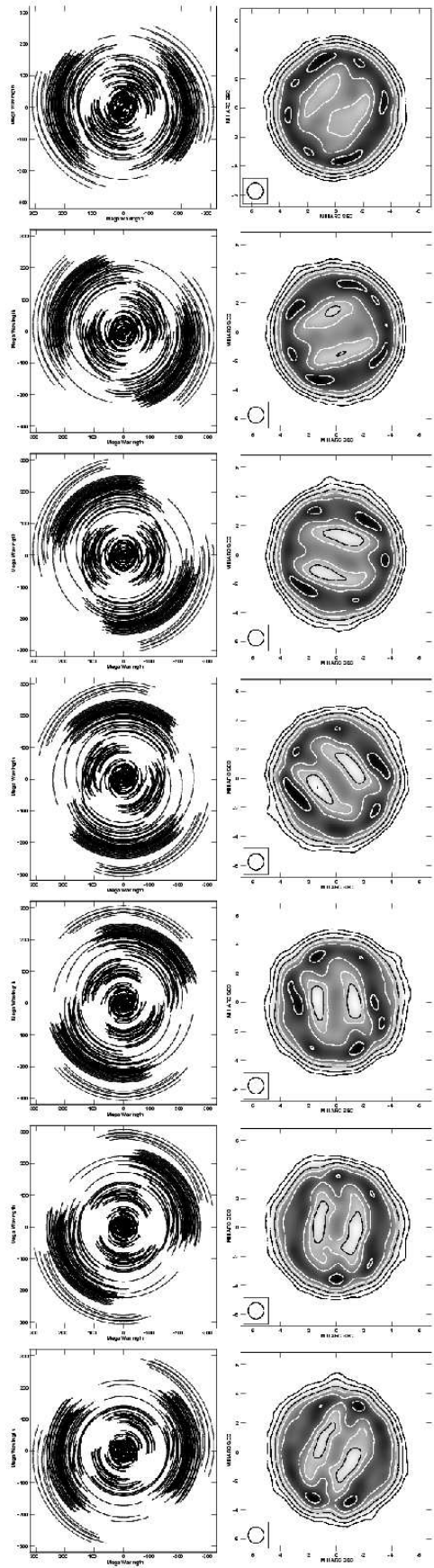}
\caption{Simulated continuous six hour VLBI observations of the model shell corresponding to day 2787, for seven different hour-angle ranges. The hour-angle coverage for the top row is 00:00 -- 06:00
and with each row the start and end time shift forwards by two hours.}
\label{fig:sixhours}
\end{figure*}

As a simplified illustration of the effects of incomplete $uv$-sampling on azimuthal structure, Figure \ref{fig:sixhours} shows a 
simulated continuous six-hour VLBI simulation for seven different hour-angle ranges. Note that the tracks are truly continuous; there are no
gaps in the visibilities. These simulations are constructed with 
an antenna configuration and input model both consistent with those of day 2787 of the actual VLBI observations of SN1993J. 
The left hand column shows the $uv$-coverage and the right hand column shows the resulting map.

The most notable feature in this series of images is that as the $uv$-coverage rotates with hour angle, \emph{so 
does the structure of the corresponding image}. The position of the peaks on the ring rotate, as does the central 
`bar' in the image. Note that the input model is purely circularly symmetric with no angular structure variations (see Figure \ref{fig:inputmodels}),
therefore \emph{none of the complex structure in the final images is real}  

\begin{figure*}
\begin{center}
\setlength{\unitlength}{1cm}
\begin{picture}(16,11.9)
\put(-1.0,-0.6){\includegraphics{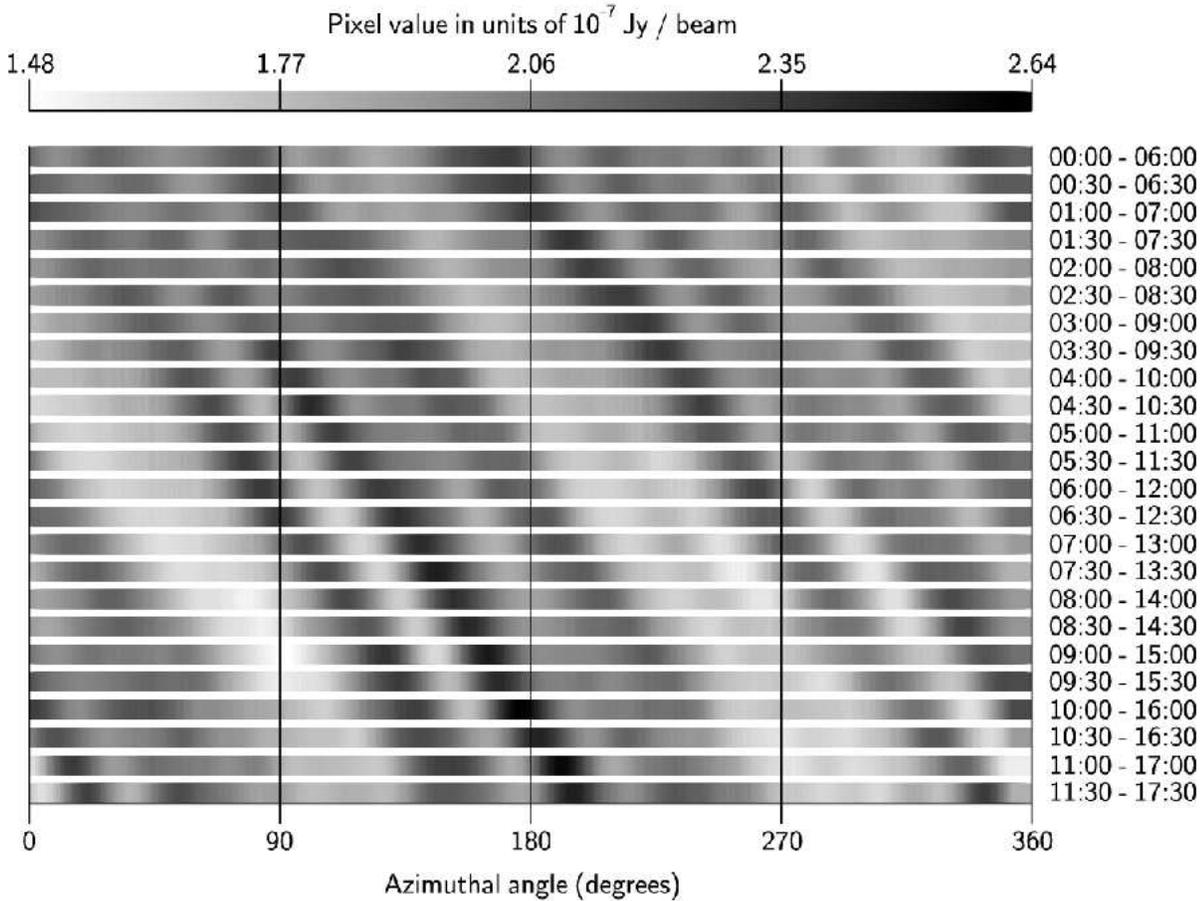}}
\end{picture}
\caption{\label{fig:greyscales}Pixel values as a function of azimuthal angle for 24 different hour angle ranges of a six-hour continuous VLBI simulation
of a source at the same high Declination as SN1993J. 
The greyscale is defined at the top of the plot and the hour angle coverage is listed adjacent to the corresponding scale strip. Pixel values are measured
at a radius of 141 pixels (3.8 milliarcseconds) from the image centre.}
\end{center}
\end{figure*}

In order to quantify the azimuthal structure that is imparted to an image, and demonstrate how it is radically affected by the 
$uv$-coverage, the pixel values as a function of angle are presented in Figure \ref{fig:greyscales}. This simulation is carried out 24 times
with the hour-angle coverage drifting by 30 minutes each time, using day 2787 as the input model. The resulting images are 512 $\times$ 512 pixels in size, and the 
pixel values are measured around a ring with a radius of 141 pixels with respect to the image centre, which on these images corresponds to 3.8 milliarcseconds. This value 
was chosen as it corresponds to the radius of the brightest annulus of emission on the input model.

Plotting the brightness distribution in this way allows one to clearly see how the structure rotates with the $uv$ coverage. 
The brightest peaks in the image appear as dark patches and the rotation manifests itself here as the diagonal striping in Figure \ref{fig:greyscales}.

\subsection{Subtle $uv$-plane changes, substantial image plane implications}

\begin{figure*}
\begin{center}
\setlength{\unitlength}{1cm}
\begin{picture}(16,9.9)
\put(0.6,-0.3){\includegraphics{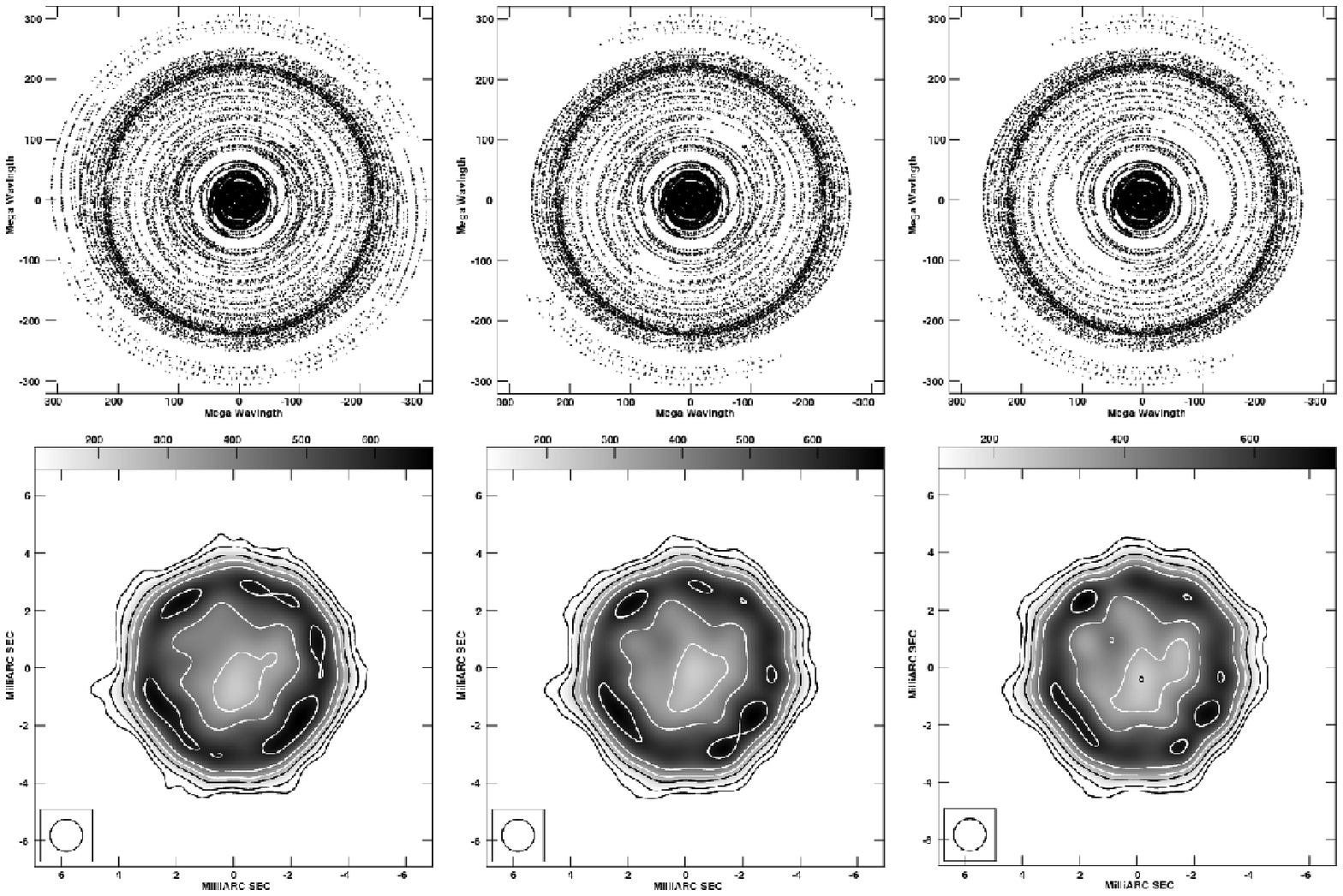}}
\end{picture}
\caption{\label{fig:uv_decay}The upper row shows the $uv$ plane coverage for day 2525 with various flagging conditions applied. 
Left to right, these are \emph{(a)} with archival
flag tables and a 15 degree antenna elevation cut off applied, \emph{(b)} 
with asymmetric Mauna Kea elevation flagging, and \emph{(c)}
with the St. Croix antenna removed.
The lower row shows the
corresponding images. The contouring regime is consistent with the simulated VLBI images, in that the greyscale runs from 16\% to 100\% of the peak brightness (listed in Table 1)
and the contour levels correspond to 1\%, 2\%, 4\%, 8\%, 16\%, 32\%, 45\%, 
64\% and 90\% of the peak brightness, starting with the first contour level above 3$\sigma$. Left to right, the peak brightnesses for these images in mJy/beam are 0.65, 0.65 and 0.66. The greyscale pixel values are in units of mJy/beam.}
\end{center}
\end{figure*}

To illustrate how apparently subtle changes to the visibility data can alter the structure in the final image, Figure \ref{fig:uv_decay} 
presents $uv$ plots and corresponding images for three different simulated observations. These are created using 
the simulation corresponding to day 2525. Contour and greyscale levels, and the restoring beam, are consistent with the day 2525 
simulation presented in Figure \ref{fig:ninemaps}.

The left hand pair shows the simulation with the 15 degree elevation cut off applied. For a source at the Declination
of SN1993J the effect this cut off has on the $uv$ coverage compared to that of a zero degree elevation cut off is minimal; in our simulation it reduced the number of visibilities by less 
than 3\%. The effect this elevation cut off has on the plot of $uv$-plane coverage is subtle and difficult to notice by eye. As such the complete plot is not presented here.

To illustrate potential problems that may be caused by local geographical features the centre pair of images simulation has an asymmetric (+35$^{\circ}$/+15$^{\circ}$) 
elevation cut off imposed on the Mauna Kea station. This drastically affects the longest baselines and consequently alters the structure in the corresponding image.
For the final image the St. Croix station is entirely removed, which mainly affects the intermediate length baselines as shown by the two prominent gaps in the $uv$ plot.

Further simulations can be found in Appendix A where we present a range of simulated images corresponding to the day 2787 epoch for a range of Declinations, 
track lengths and hour angle ranges.

\section{Conclusion}

We have demonstrated in this paper how sparse $uv$ coverage can impart apparently complex azimuthal structure to a radio brightness
distribution that is in reality morphologically simple. Rotation of the $uv$-plane sampling pattern determined by varying hour angle
coverage leads to a corresponding rotation in the radio image. 

This is just one example of the kind of structures that are vulnerable to errors arising from sparse $uv$-plane sampling. For a full
discussion of other structures and their vulnerabilities we refer the reader to the works by Cornwell (1999) and Briggs (1995) 
who investigate the use of different deconvolution methods which further undermine these structures and cannot completely overcome the problems caused
by incomplete information.

By reconstructing the $uv$-plane distributions achieved in the SN1993J VLBI campaign and convolving these with a perfect
spherically symmetric model of the supernova, we have shown that sparse $uv$ coverage may be wholly or partially responsible
for the evolving complex azimuthal structure observed in real supernova remnant.

\section*{Acknowledgements}

The authors wish to thank Amy Mioduszewski for kindly allowing us to use her synchrotron models.  We thank the referees for their 
comments, and for highlighting some potential issues with UVCON. We also wish to thank Leonia Kogan and Eric Greisen of NRAO
for rapidly addressing these issues. 
IH thanks the Science and Technology Facilities Council. 
KMB thanks the Royal Society for a University Research Fellowship. IH and KMB are very grateful to the Leverhulme Trust whose prize supported this work. 
HRK thanks the University of Oxford for financial support.

\appendix

\section{Track length and Declination dependence}

The parameter space defined by factors which will affect image fidelity is enormous. Both the realistic and illustrative simulations presented
in this paper represent a very specific case in terms of source Declination, track length and therefore $uv$-plane coverage.

\begin{figure*}
\begin{center}
\setlength{\unitlength}{1cm}
\begin{picture}(16,22)
\put(-0.3,-0.5){\includegraphics{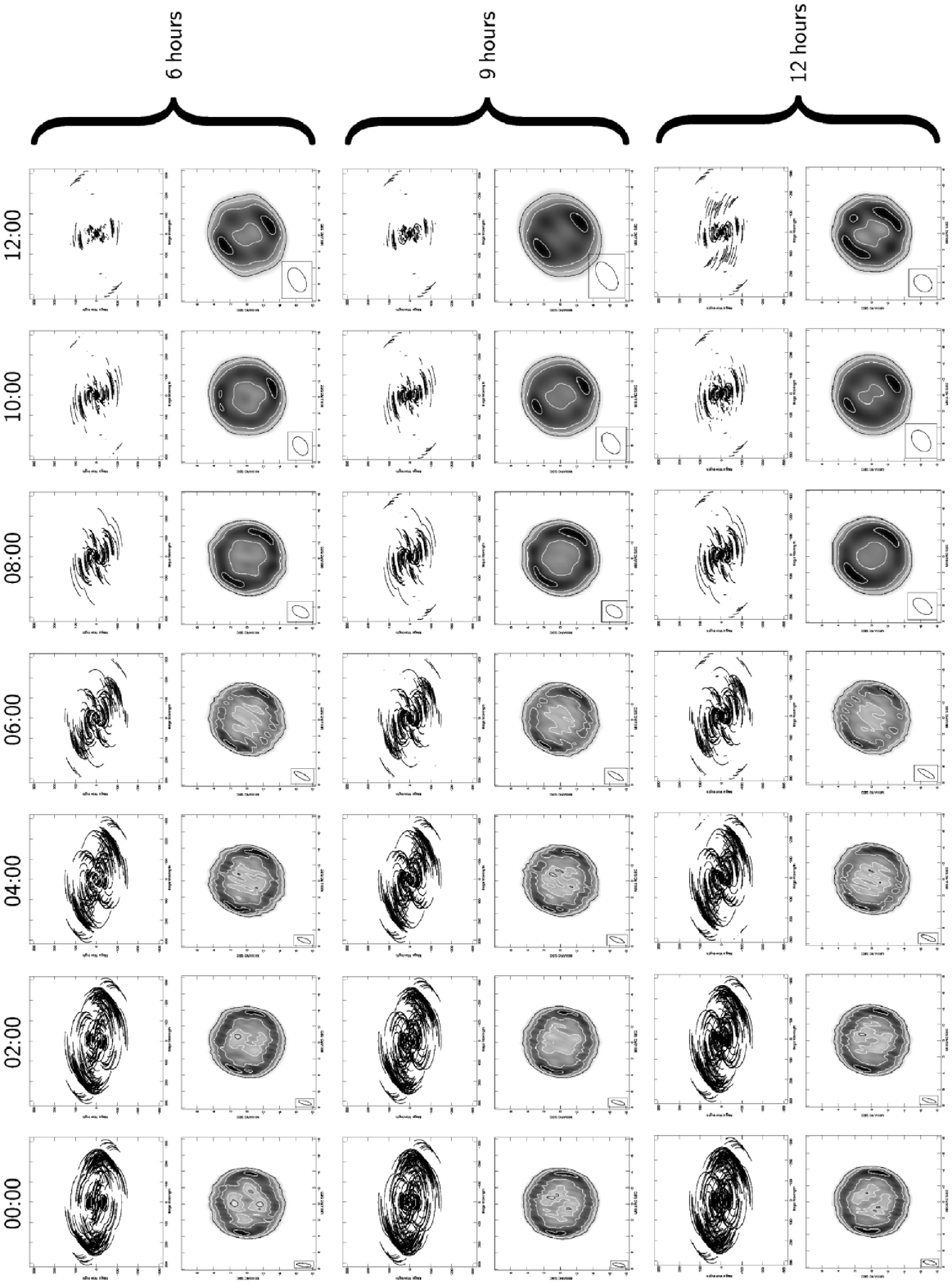}}
\end{picture}
\caption{\label{fig:d20}Simulations of the day 2787 model at 20 degrees Declination.}
\end{center}
\end{figure*}

\begin{figure*}
\begin{center}
\setlength{\unitlength}{1cm}
\begin{picture}(16,22)
\put(-0.3,-0.5){\includegraphics{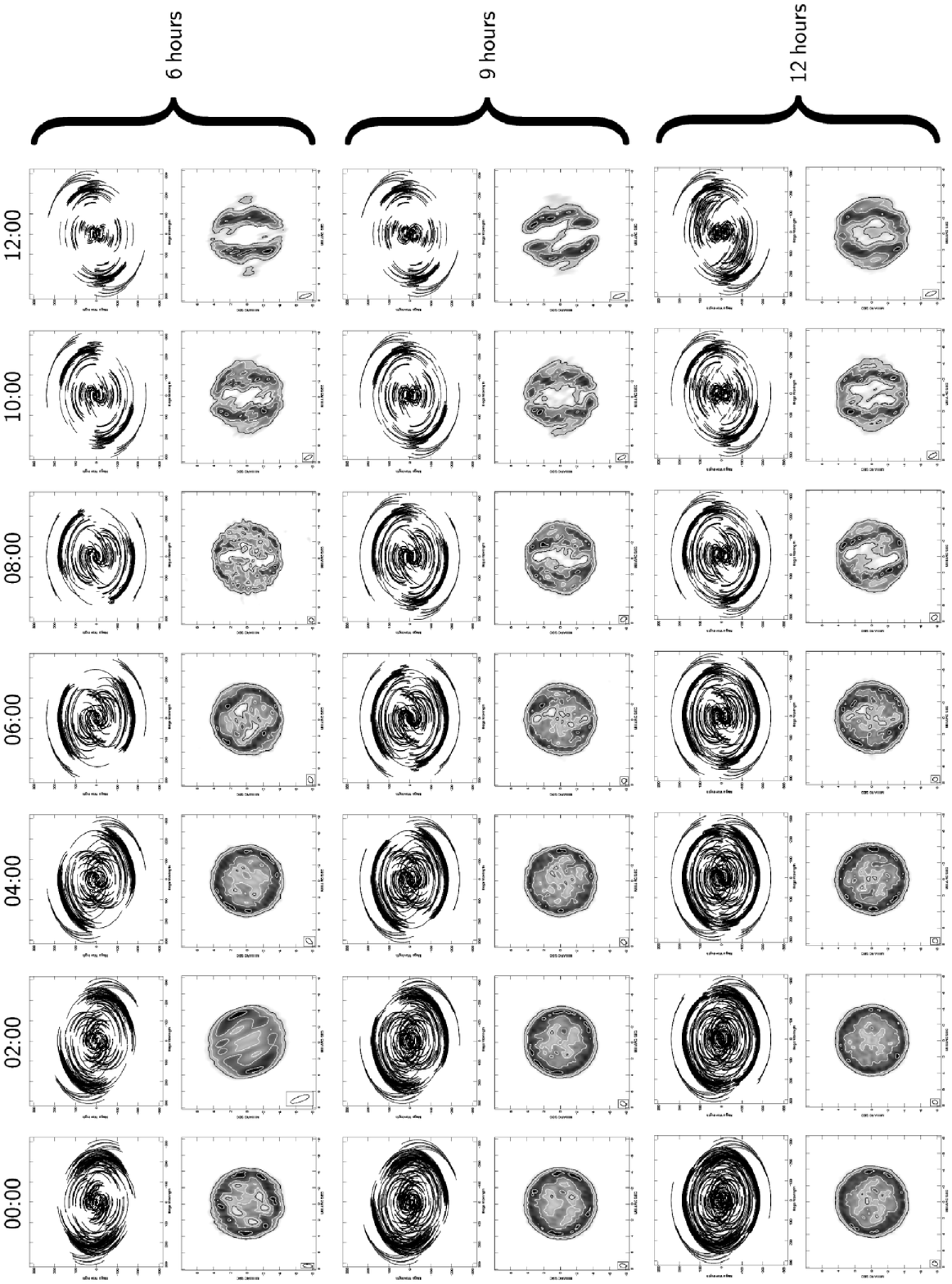}}
\end{picture}
\caption{\label{fig:d40}Simulations of the day 2787 model at 40 degrees Declination.}
\end{center}
\end{figure*}

\begin{figure*}
\begin{center}
\setlength{\unitlength}{1cm}
\begin{picture}(16,22)
\put(-0.3,-0.5){\includegraphics{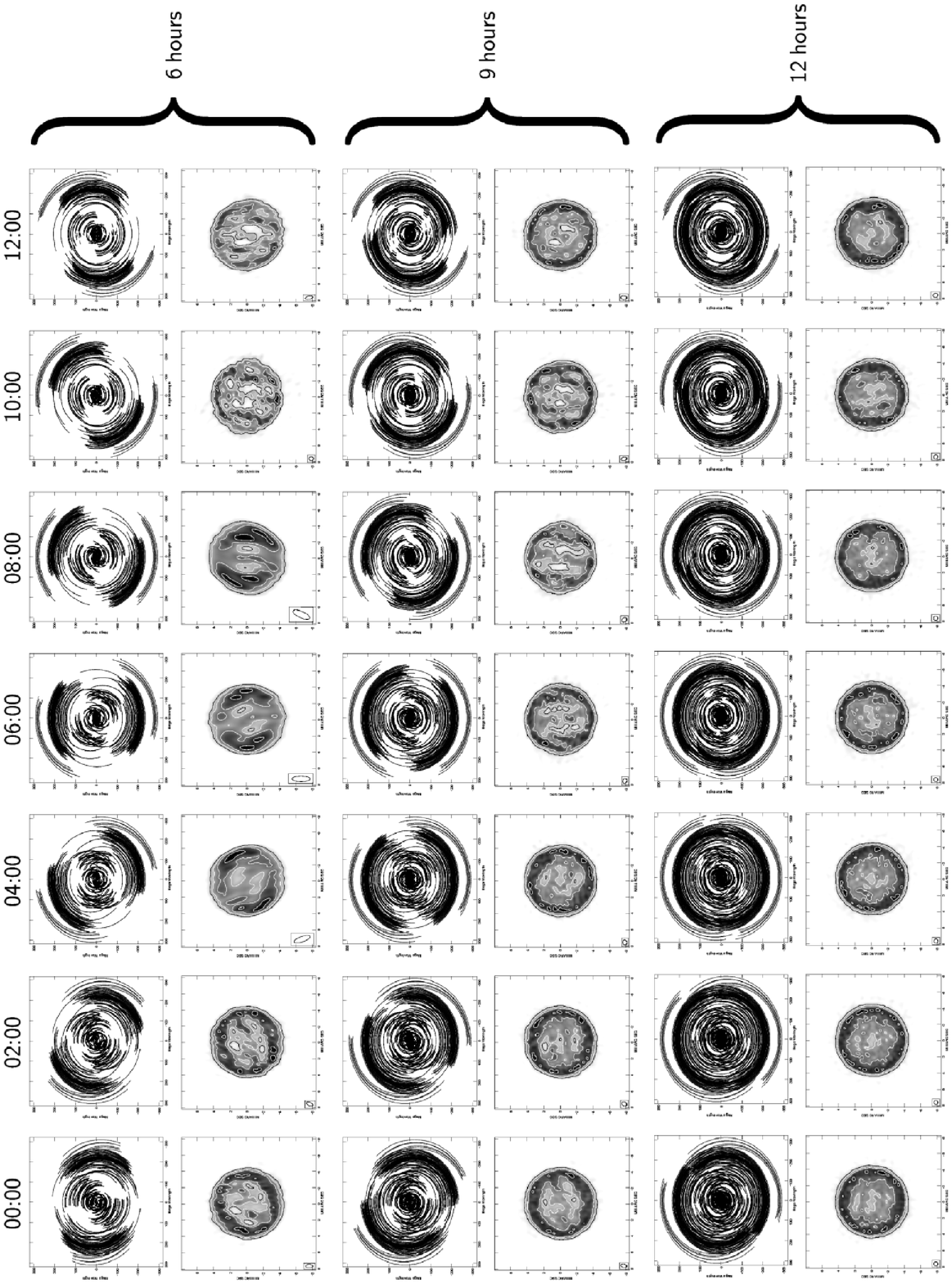}}
\end{picture}
\caption{\label{fig:d60}Simulations of the day 2787 model at 60 degrees Declination.}
\end{center}
\end{figure*}

\begin{figure*}
\begin{center}
\setlength{\unitlength}{1cm}
\begin{picture}(16,22)
\put(-0.3,-0.5){\includegraphics{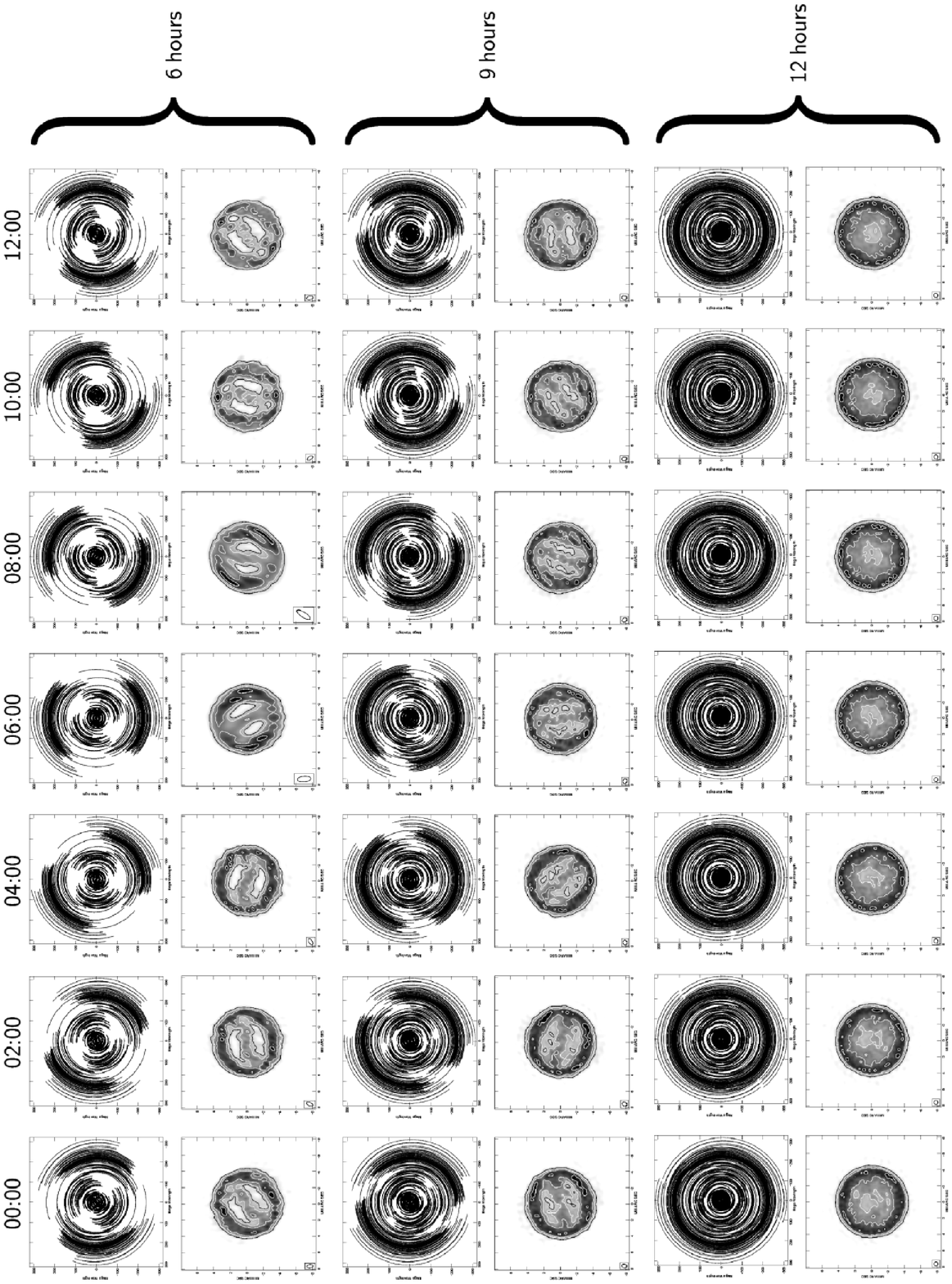}}
\end{picture}
\caption{\label{fig:d80}Simulations of the day 2787 model at 80 degrees Declination.}
\end{center}
\end{figure*}

In this section we present simulations of the day 2846 model at Declinations of 20, 40, 60 and 80 degrees with 6, 9 and 12 hour continuous observations.
Please refer to Figure \ref{fig:inputmodels} to see the morphology of the input source.
Seven unique hour-angle ranges are simulated for each declination-track length pair resulting in 84 unique observations. Simulated images are grouped by
Declination and are presented in Figures \ref{fig:d20}, \ref{fig:d40}, \ref{fig:d60} and \ref{fig:d80}. 

The greyscale range in each image is from 16\% to 100\% of the peak value and the contours are 32\%, 45\%, 64\% and 90\% 
of the peak flux. The starting value of the hour-angle range is listed on the left hand side of the figures, with the end hour angle being the start
value plus the track length, which is printed at the top of each page. 

Imaging of the simulated data sets was automated in this case. Cleaning was not carried out interactively and the clean cycle was terminated after reaching
1000 components. The Gaussian fitted to the half power point of the central lobe of the point spread function for each image was used as the restoring beam
in each case. This is plotted in the lower left of each figure. The favourable Declination (+69 degrees) of SN1993J facilitates the use of a circular 
restoring beam as the fitted beam itself is nearly circular. For lower Declinations this is not applicable, as can be easily seen by the elongated beams
of lower Declination simulations. 

Note the wide variety of image artifacts that can arise due to the $uv$-plane sampling. Very sparse coverage at low Declinations 
leads to `polar cap'-type features, striped gaps in the $uv$-coverage result in dual ridge features, and central bars across the shell are 
a regular outcome. Only the 12-hour tracks at high Declination, featuring almost complete coverage, result in a ring structure with a central minimum. Artificial azimuthal structure is present
in each of these simulated images, even those with excellent $uv$-coverage.

\bsp 

\label{lastpage}

\end{document}